\documentclass[
    ,final            
  ,numberedheadings 
  ]
  {aipproc}

\layoutstyle{6x9}

\begin{document}

\title{Hadronic physics at KLOE}



\author{Roberto Versaci on behalf of the KLOE collaboration
\footnote{  F.~Ambrosino,  A.~Antonelli,  M.~Antonelli,  F.~Archilli,
  P.~Beltrame,  G.~Bencivenni,  S.~Bertolucci,  C.~Bini,  C.~Bloise,
  S.~Bocchetta,  F.~Bossi,  P.~Branchini,  G.~Capon,  T.~Capussela,
  F.~Ceradini,  P.~Ciambrone,  E.~De~Lucia,  A.~De~Santis,  P.~De~Simone,
  G.~De~Zorzi,  A.~Denig,  A.~Di~Domenico,  C.~Di~Donato,  B.~Di~Micco,
  M.~Dreucci,  G.~Felici,  S.~Fiore,  P.~Franzini,  C.~Gatti,  P.~Gauzzi,
  S.~Giovannella,  E.~Graziani,  G.~Lanfranchi,  J.~Lee-Franzini,
  M.~Martini,  P.~Massarotti,  S.~Meola,  S.~Miscetti,  M.~Moulson,
  S.~M\"uller,  F.~Murtas,  M.~Napolitano,  F.~Nguyen,  M.~Palutan,
  E.~Pasqualucci,  A.~Passeri,  V.~Patera,  P.~Santangelo,  B.~Sciascia,
  T.~Spadaro,  M.~Testa,  L.~Tortora,  P.~Valente,  G.~Venanzoni,
  R.Versaci,  G.~Xu}}{
  address={Dipartimento di Energetica dell'Universit\`a ``Sapienza'',
           Roma, Italy},
  altaddress={Laboratori Nazionali di Frascati dell'INFN, Frascati, Italy}
}

\begin{abstract}
  New KLOE results on scalar mesons, $\gamma\gamma$ physics and $\eta$
  physics are presented.
\end{abstract}

\maketitle


\section{Scalar physics}

  The structure of the scalars below 1 GeV needs to be further clarified.
  Several models have been proposed to describe them (e.g. $q\overline{q}$, 
  four quarks, $K\overline{K}$ molecules, etc.).
  The decay of the scalars into two pseudoscalars ($S \to P P'$) 
  can be used to investigate their nature because the branching ratios and 
  the invariant mass of the two pseudoscalars are sensitive to the scalar 
  structure.

  \paragraph{$\phi \to a_0(980) \gamma \to \eta \pi^0 \gamma$ 
    decay \cite{a0}}

  For this measurement about 400 pb$^{-1}$ of KLOE collected data have been
  used.
  The analysis has been performed for two different $\eta$ final states, 
  i.e. $\eta \to \gamma \gamma$ and $\eta \to \pi^+ \pi^- \pi^0$.
  A kinematic fit has been performed imposing the four momentum 
  conservation, the photon velocity and the invariant masses of both $\eta$
  and $\pi^0$.
  The $\eta \pi^0$ invariant mass distribution has been fitted with the
  ``no-structure''\cite{nostructure} and the ``kaon loop''\cite{kaonloop}
  models after background subtraction.
  The results of the fit are shown in table \ref{tab:a0}.
  It is interesting to note that both models give a large coupling 
  of the $a_0(980)$ with the $\phi$ meson, indicating a sizable 
  strange quark content in the $a_0(980)$.
  The branching ratio obtained for the two different decay chains are
  in agreement:
  $BR(\phi\to\eta\pi^0\gamma)=(7.01\pm0.10_{stat.}\pm0.20_{syst.})\times 10^{-5}$
  for the $\eta \to \gamma \gamma$ final state and 
  $BR(\phi\to\eta\pi^0\gamma)=(7.12\pm0.13_{stat.}\pm0.22_{syst.})\times 10^{-5}$
  for the $\eta \to \pi^+ \pi^- \pi^0$ final state.
  
  \begin{table}
    \begin{tabular}{lll}
      \hline
      & \tablehead{1}{l}{b}{Kaon loop}
      & \tablehead{1}{l}{b}{No structure} \\
      \hline
      $M_{a_0}$ [MeV]              
      & $982.5\ \pm\ 1.6\ \pm\ 1.1$  
      & $982.5$ (Fixed) \\
      $g_{a_0 K^+ K^-}$ [GeV]         
      & $2.15\ \pm\ 0.06\ \pm\ 0.06$ 
      & $2.01\ \pm\ 0.07\ \pm\ 0.28$ \\
      $g_{a_0 \eta \pi^0}$ [GeV]       
      & $2.82\ \pm\ 0.03\ \pm\ 0.04$ 
      & $2.46\ \pm\ 0.08\ \pm\ 0.11$ \\
      $g_{\phi a_0 \gamma}$ [GeV$^{-1}$] 
      & $1.58\ \pm\ 0.10\ \pm\ 0.16$\tablenote{Not a free parameter of the
        fit in this model. Calculated from other fit outputs.}
      & $1.83\ \pm\ 0.03\ \pm\ 0.08$ \\
      BR($\phi\to\rho\pi\to\eta\pi\gamma$) 
      & $(0.92\ \pm\ 0.40\ \pm\ 0.15)\cdot 10^{-6}$ 
      & $(0.05\ \pm\ 4\ \pm\ 0.07)\cdot 10^{-6}$ \\
      BR($\eta\to\gamma\gamma$)/BR($\eta\to\pi^+\pi^-\pi^0$)
      & $1.70\ \pm\ 0.04\ \pm\ 0.03$
      & $1.70\ \pm\ 0.03\ \pm\ 0.01$\\
      $\chi^2$ probability        
      & 0.104 
      & 0. 309\\
      \hline
    \end{tabular}
    \caption{Output of the fit to the $\eta\pi^0$ invariant mass with 
      two different models: kaon loop and no structure.}
    \label{tab:a0}
  \end{table}

  \paragraph{$\phi \to K^0 \bar{K^0} \gamma$ decay \cite{k0k0g}}
  
\begin{figure}[b!]
    \centering
    \begin{minipage}[c]{0.4\linewidth}
      \begin{center}
        \includegraphics[height=.2\textheight]{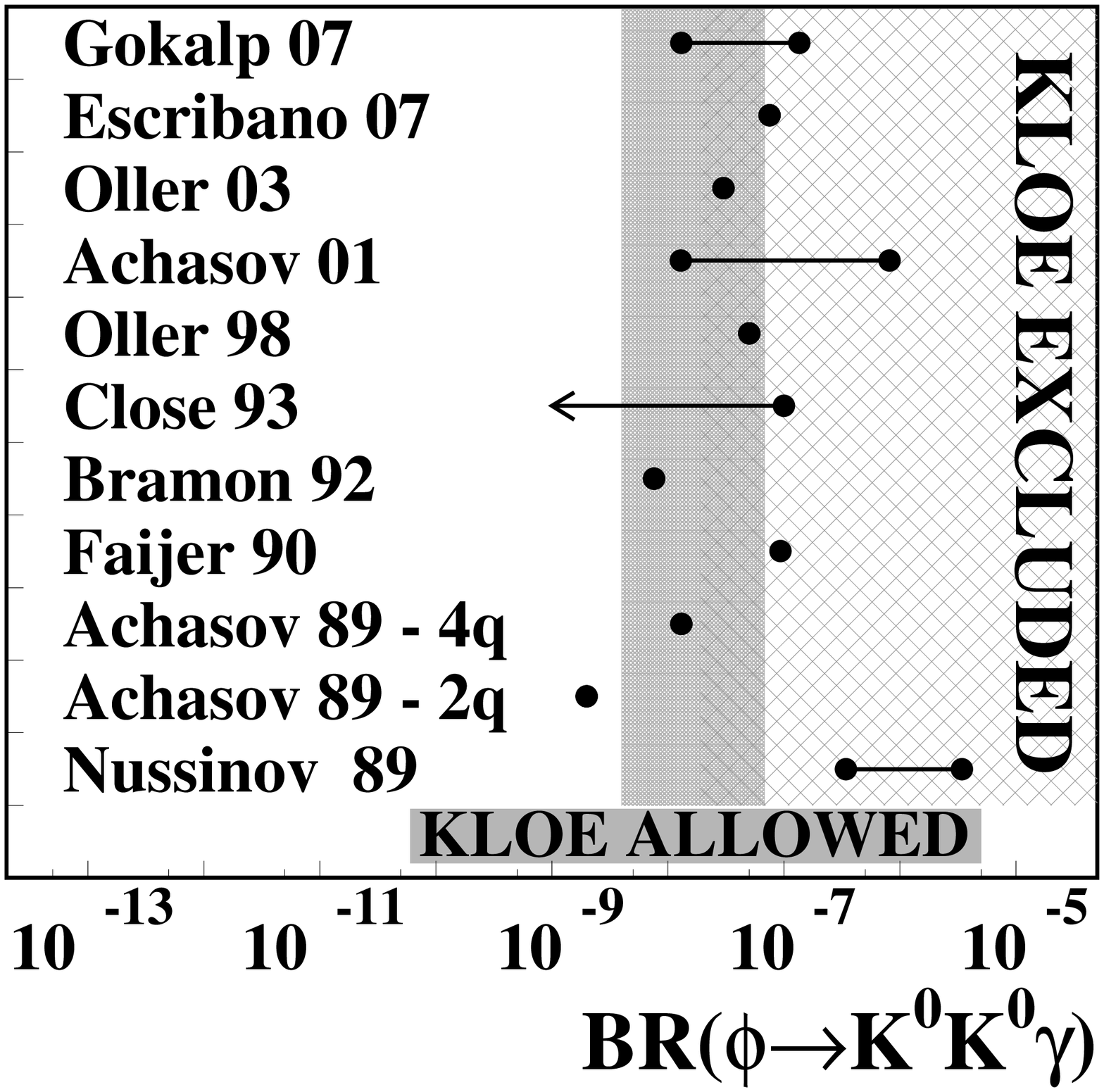}
      \end{center}
    \end{minipage}
    \begin{minipage}[c]{0.6\linewidth}
      \begin{center}
        \includegraphics[height=.175\textheight]{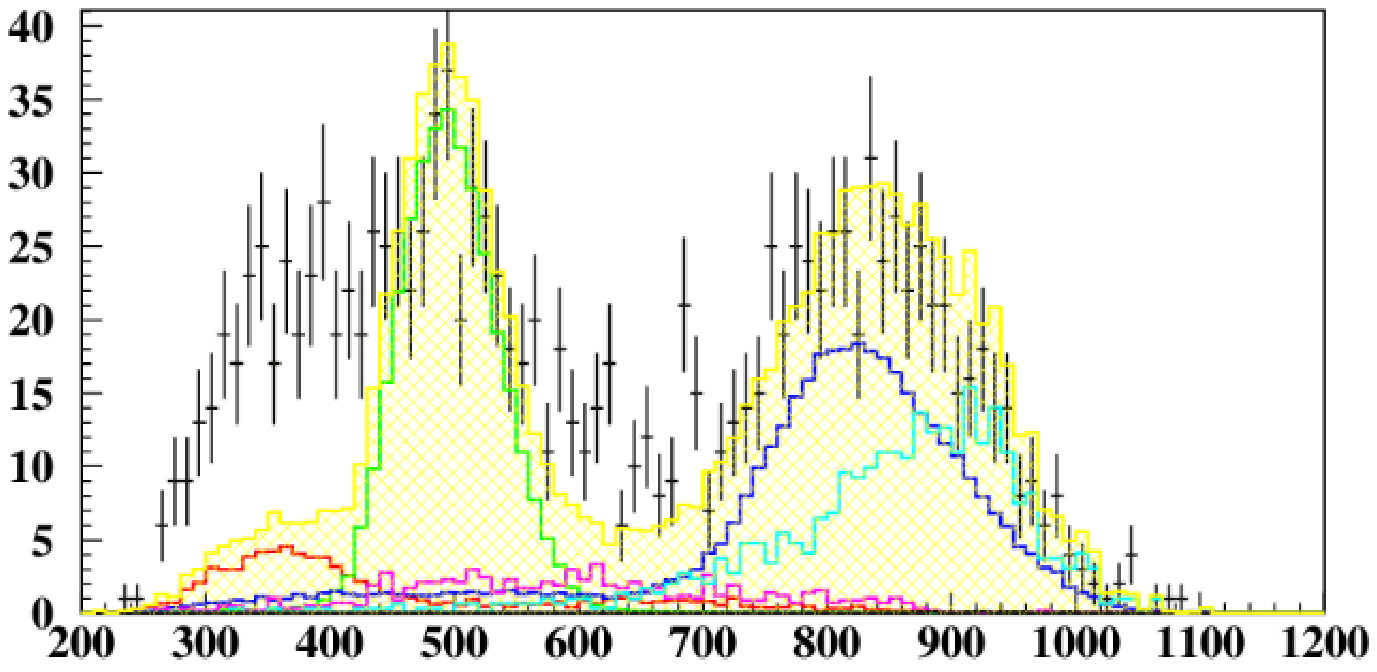}\\
        M$_{4\gamma}$ [MeV]
        \caption{\textbf{Left}: 
          comparison between theoretical predictions for the 
          $\phi \to K^0 \bar{K^0} \gamma$ branching ratio and KLOE upper
          limit. 
          The grey area represents the expected range for the branching ratio
          using other KLOE results on scalar mesons.
          \textbf{Right}: search for
          $\gamma\gamma\to\sigma(600)\to\pi^0\pi^0$; 
          fit to the invariant mass of the four photons, M$_{4\gamma}$,
          with MC shapes of expected source of background. 
          Dots: data; atched yellow: total background; 
          red: $\phi \to \eta\gamma \to \pi^0\pi^0\pi^0\gamma$;
          blue: $e^+e^- \to \omega\pi^0 \to \pi^0\pi^0\gamma$;
          green: $\phi \to K_S K_L$; cyan: $\phi \to f_0\gamma$;
          magenta: $e^+e^- \to \gamma\gamma$.}
        \label{fig:k0k0g-gg}
      \end{center}
    \end{minipage}
  \end{figure}

  This decay allegedly proceeds through the intermediate $f_0(980)$ (I=0)
  and $a_0(980)$ (I=1) scalar mesons: 
  $\phi \to (f_0 + a_0) \gamma \to K^0 \bar{K^0} \gamma$.
  The kaon pair is produced in a $J^{PC} = 0^{++}$ state, so the two kaons
  are both $K_S$ or $K_L$.
  We have searched for a final state with a $K_S K_S$, with both $K_S$' 
  decaying to $\pi^+ \pi^-$.
  This request reduces the probability of observation to $\sim 22\%$, but
  selects a class of event with a clear signature: 
  four tracks and a low energy photon coming from the interaction point.
  In this analysis the whole KLOE dataset, $\sim 2.2$ fb$^{-1}$, has been
  used. 
  At the end of which we have observed 5 events in the data, while
  we were expecting $3.2\pm0.7$ background events from MC.
  A Cousin-Feldman approach has been used \cite{feldmancousins} and a 90\%
  confidence level upper limit on the branching ratio has been obtained:
  $BR(\phi \to K^0 \bar{K^0} \gamma) < 1.9 \times 10^{-8}$.
  This measurement excludes some of the theoretical predictions and is in
  agreement with expectations from other KLOE measurements
  (see figure \ref{fig:k0k0g-gg} left, reference \cite{k0k0g} and
  references therein).
  
  \section{$\gamma \gamma$ physics}
  
  KLOE has been making a pilot study for the search
  $\gamma\gamma\to\sigma(600)\to\pi^0\pi^0$ using 11 pb$^{-1}$ from the 
  240 pb$^{-1}$ taken at $\sqrt{s}=1000$ MeV \cite{gammagamma}.
  At this energy the background from $\phi$ decays is very small. 
  We have performed a fit to the four photons invariant mass
  (M$_{4\gamma}$), using the shapes of the known sources,
  see figure \ref{fig:k0k0g-gg}, right. 
  The result of the fit is very poor $\chi^2$/dof = 441/94, showing an 
  excess of events in the expected $\sigma(600)$ region, compared to what
  expected from MC, therefore pointing towards a search for the signal in
  the 240 pb$^{-1}$.
  
\section{Pseudoscalar physics}

  The $\phi$ meson decays about 1.3\% of times into $\eta\gamma$, this
  implies DA$\Phi$NE is an $\eta$-factory.
  KLOE has collected one of the largest sample of $\eta$ mesons in the
  world, about $10^8$.

  \paragraph{$\eta-\eta'$ mixing and $\eta'$ gluonium content \cite{etaetap}}

    The KLOE paper on $\eta-\eta'$ mixing \cite{etaetap}, suggesting
    for a $3\sigma$ evidence of gluonium content in the $\eta'$ meson,
    has triggered a large amount of discussion among theoreticians.
    Therefore we have decided to perform a new and more detailed study of 
    this topic.
    We have considered $\eta$ and $\eta'$ in the quark mixing base as 
    described in \cite{rosner}
    ($|\eta'> = X_{\eta'}|q\bar{q}> +  Y_{\eta'}|s\bar{s}> + Z_G|G>$).
    The new fit we have performed has more constraints thus allowing an 
    independent determination of more free parameters.
    We use the BR values from PDG 2008 \cite{pdg2008} and the new KLOE
    results on the $\omega$ meson \cite{omega}.
    The fit has been performed both imposing the gluonium content to be
    zero or allowing it free.
    The results are shown in table \ref{tab:gluonium}: gluonium content of
    the $\eta'$ is confirmed at $3\sigma$ level.
    \begin{table}
       \begin{tabular}{lll}
          \hline
          & \tablehead{1}{l}{b}{Gluonium content forced to be zero}
          & \tablehead{1}{l}{b}{Gluonium content free} \\
          \hline
          $Z_G^2$        & fixed 0             & $0.115\ \pm\ 0.036$ \\
          $\phi_P$       & $(41.4\ \pm\ 0.5)^\circ$  & $(40.4\ \pm\ 0.6)^\circ$ \\
          $Z_q$          & $0.93\ \pm\ 0.02$   & $0.94\ \pm\ 0.03$   \\
          $Z_s$          & $0.82\ \pm\ 0.05$   & $0.83\ \pm\ 0.05$   \\
          $\phi_V$       & $(3.34\ \pm\ 0.09)^\circ$ & $(3.32\ \pm\ 0.09)^\circ$\\
          $m_s/\bar{m}$  & $1.24\ \pm\ 0.07$   & $1.24\ \pm\ 0.07$   \\
          $\chi^2$ / dof & 14.7/4              & 4.6/3               \\
          P($\chi^2$)    & 0.005               & 0.20                \\
          \hline
       \end{tabular}
       \caption{Output of the fit imposing or not the gluonium content to 
         be zero.}
       \label{tab:gluonium}
    \end{table}

  \paragraph{$\eta$ decays into four charged particles \cite{etappee}}

\begin{figure}
      \centering
      \begin{minipage}[c]{0.3\linewidth}
        \begin{center}
          \includegraphics[height=.2\textheight]{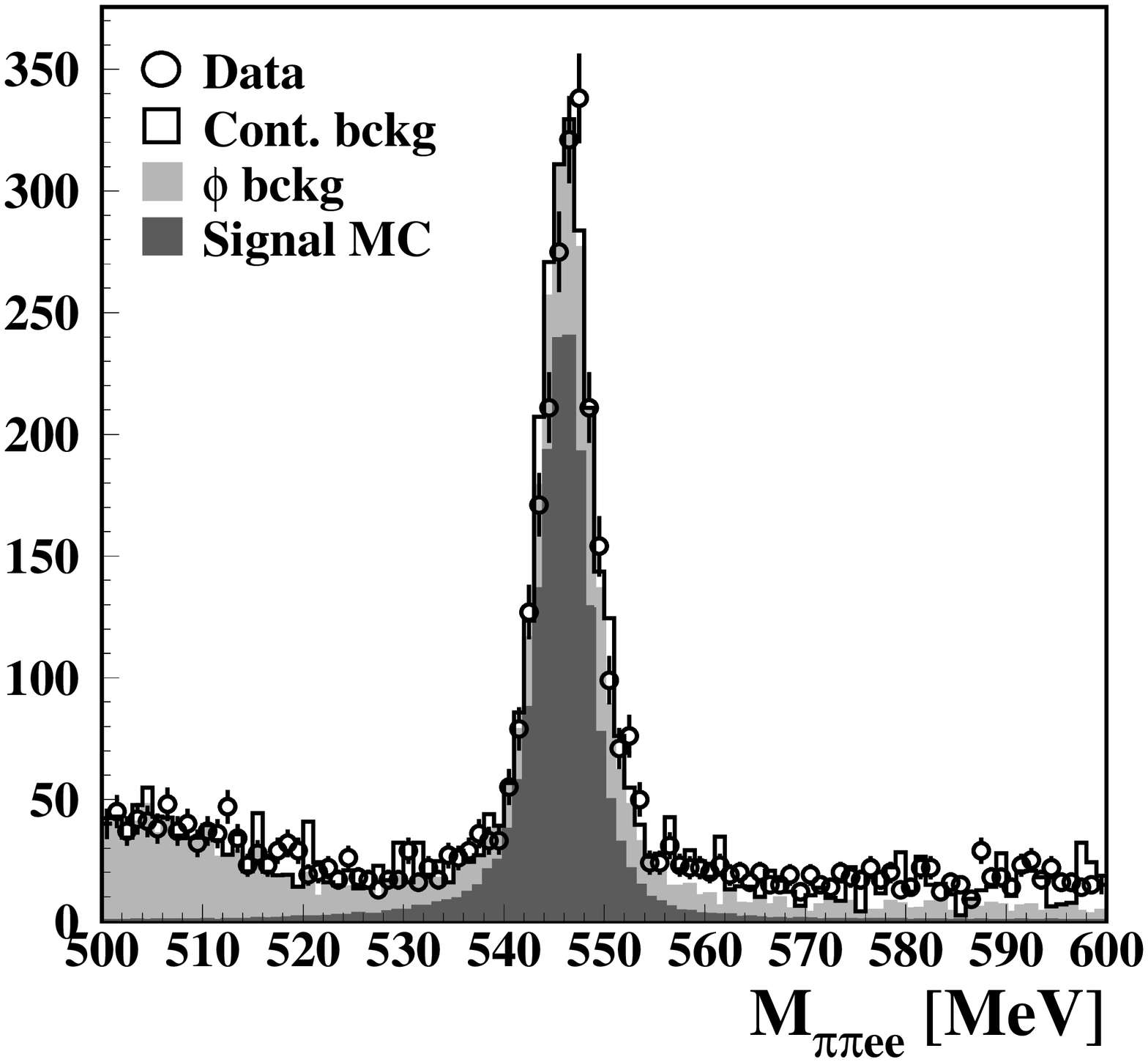}
        \end{center}
      \end{minipage}
      \hspace{0.5cm}
      \begin{minipage}[c]{0.3\linewidth}
        \begin{center}
          \includegraphics[height=.2\textheight]{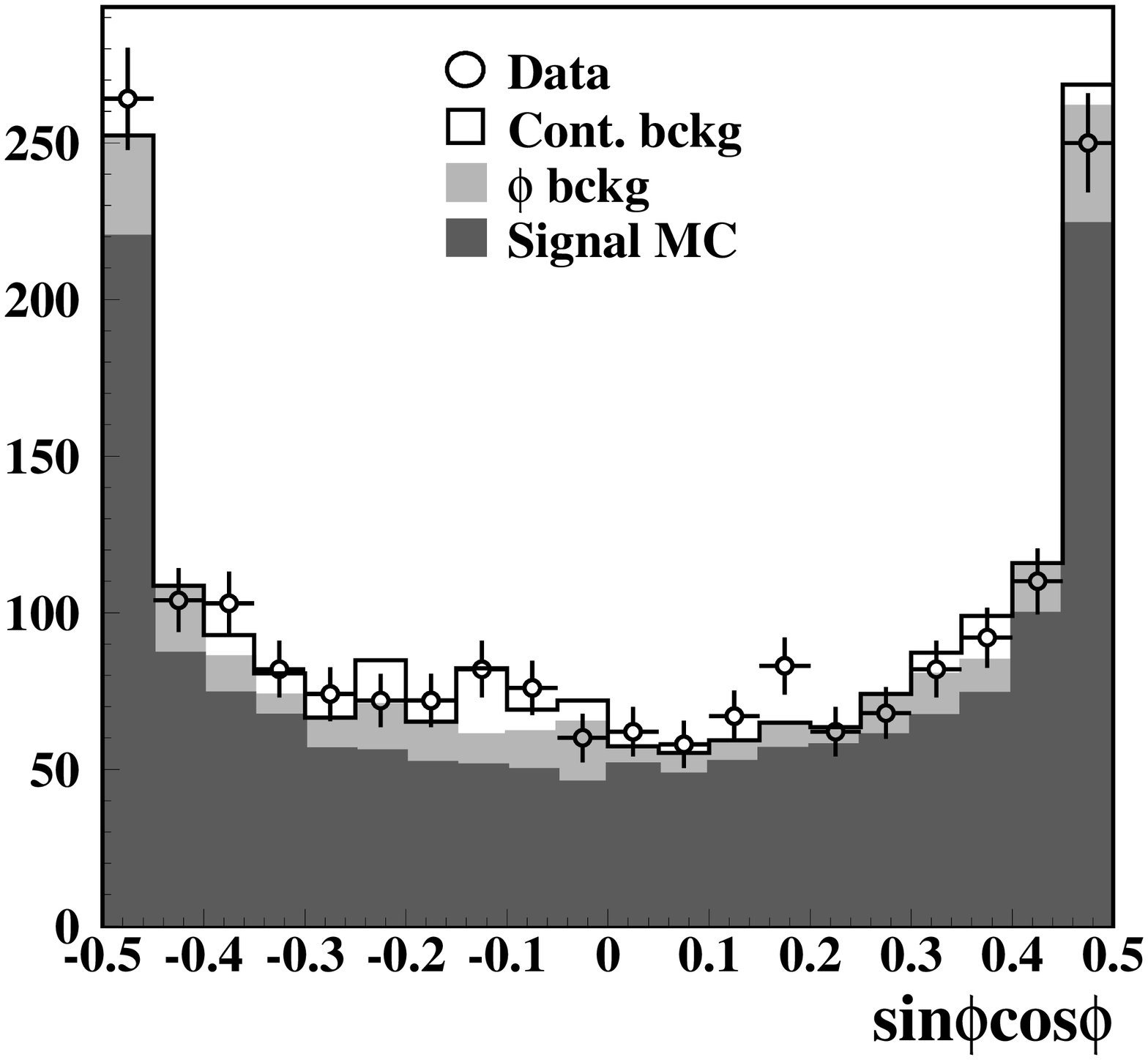}
        \end{center}
      \end{minipage}
      \hspace{0.5cm}
      \begin{minipage}[c]{0.3\linewidth}
        \begin{center}
          \vspace{0.5cm}
          \includegraphics[height=.2\textheight]{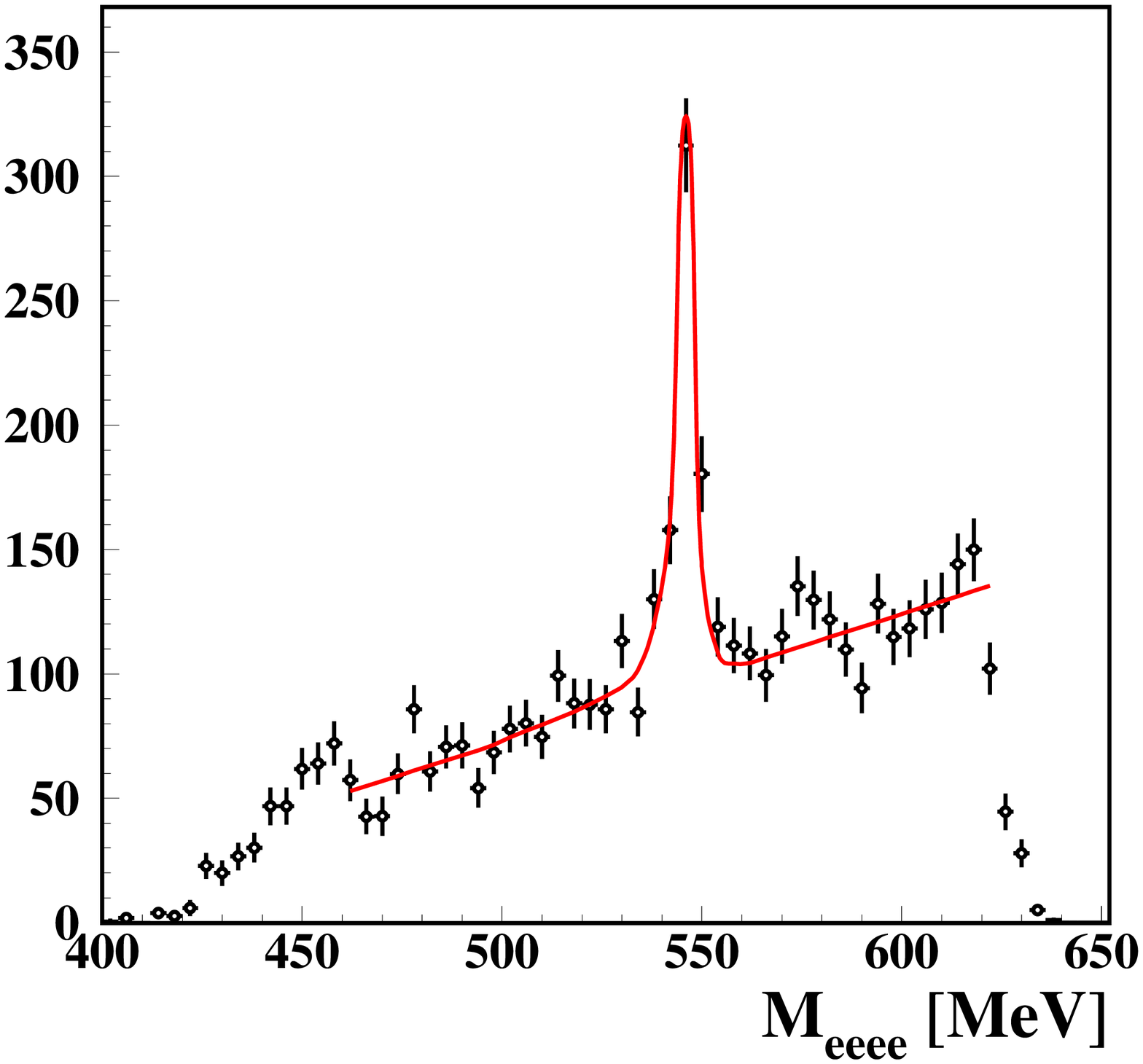}
          \caption{\textbf{Left and center}: 
            $\eta \to \pi^+\pi^-e^+e^-$ analysis;
            $\pi^+\pi^-e^+e^-$ invariant mass and angular asymmetry
            distributions.
            Dots: data. 
            The black histogram is the expected distribution, 
            i.e. signal MC (dark grey), 
            $\phi$ background (light grey) and continuum background (white).
            \textbf{Right}:
            $\eta\to e^+ e^- e^+ e^-$ analysis;
            fit of the four electron invariant mass, M$_{eeee}$.
          } 
          \label{fig:eta4c}
        \end{center}
      \end{minipage}
    \end{figure}

    KLOE has started to study the decays of the $\eta$ into four charged
    particles, using 1.7 fb$^{-1}$ of data.
    This decay is interesting because it allows us to probe the $\eta$
    internal structure exploiting the conversion of the virtual photon into
    a lepton pair \cite{landsberg}.
    It is also interesting because a non-CKM CP violating mechanism has
    been suggest to be present in this decay\cite{gao}, and should manifest
    as an angular asymmetry $A_\phi$, between the pion and the electron
    decay planes in the $\eta$ rest frame. 
    After background rejection a fit of the sidebands of the four tracks 
    invariant distribution has been performed to obtain the background
    scale factors.
    Most of the background is due to $\phi$ decays, but there is still a
    non-negligible contribution from continuum events.
    Signal events have been counted in the $\eta$ mass region, giving
    $BR(\eta\to\pi\pi ee)=(26.8\pm0.9_{Stat.}\pm0.7_{Syst.})\times 10^{-5}$
    and $A_\phi=(-0.6\pm2.5_{Stat.}\pm1.8_{Syst.})\times 10^{-2}$
    \cite{etappee}, see figure \ref{fig:eta4c} left and center.

    More recently KLOE has started studying the $\eta\to e^+ e^- e^+ e^-$ 
    decay.
    This decay, together with the $\eta\to \mu^+ \mu^- e^+ e^-$, is 
    interesting for the $\eta$ meson form factor because there
    are only leptons in the final state.
    The analysis strategy is similar to the $\pi\pi ee$ one.
    Most of the background comes from continuum events and a small
    contribution is due to $\phi$ decays.
    The latter is subtracted from data using the MC shape.
    The number of events is obtained fitting the data distribution of the 
    4 electron invariant mass, M$_{eeee}$, with signal and background
    shapes (figure \ref{fig:eta4c}, right).
    From the fit we obtain 413 events.
    This constitutes the first observation of this decay.

\bibliographystyle{aipproc}   
\bibliography{cipanp}

\hyphenation{Post-Script Sprin-ger}
\begin{thebibliography}{13}
\expandafter\ifx\csname natexlab\endcsname\relax\def\natexlab#1{#1}\fi
\providecommand{\enquote}[1]{``#1''}
\expandafter\ifx\csname url\endcsname\relax
  \def\url#1{\texttt{#1}}\fi
\expandafter\ifx\csname urlprefix\endcsname\relax\def\urlprefix{URL }\fi
\providecommand{\eprint}[2][]{\url{#2}}

\bibitem[Ambrosino et~al.(2009{\natexlab{a}})]{a0}
F.~Ambrosino, et~al.  (2009{\natexlab{a}}), \eprint{0904.2539}.

\bibitem[Isidori et~al.(2006)]{nostructure}
G.~Isidori, L.~Maiani, M.~Nicolaci, and S.~Pacetti, \emph{JHEP} \textbf{05},
  049 (2006), \eprint{hep-ph/0603241}.

\bibitem[Achasov and Kiselev(2003)]{kaonloop}
N.~N. Achasov, and A.~V. Kiselev, \emph{Phys. Rev.} \textbf{D68}, 014006
  (2003), \eprint{hep-ph/0212153}.

\bibitem[Ambrosino et~al.(2009{\natexlab{b}})]{k0k0g}
F.~Ambrosino, et~al.  (2009{\natexlab{b}}), \eprint{0903.4115}.

\bibitem[Feldman and Cousins(1998)]{feldmancousins}
G.~J. Feldman, and R.~D. Cousins, \emph{Phys. Rev.} \textbf{D57}, 3873--3889
  (1998), \eprint{physics/9711021}.

\bibitem[Ambrosino et~al.(2008{\natexlab{a}})]{gammagamma}
F.~Ambrosino, et~al., \emph{Nuovo Cim.} \textbf{31C}, 415 (2008{\natexlab{a}}).

\bibitem[Ambrosino et~al.(2007)]{etaetap}
F.~Ambrosino, et~al., \emph{Phys. Lett.} \textbf{B648}, 267--273 (2007),
  \eprint{hep-ex/0612029}.

\bibitem[Rosner(1983)]{rosner}
J.~L. Rosner, \emph{Phys. Rev.} \textbf{D27}, 1101 (1983).

\bibitem[Amsler et~al.(2008)]{pdg2008}
C.~Amsler, et~al., \emph{Phys. Lett.} \textbf{B667}, 1 (2008).

\bibitem[Ambrosino et~al.(2008{\natexlab{b}})]{omega}
F.~Ambrosino, et~al., \emph{Phys. Lett.} \textbf{B669}, 223--228
  (2008{\natexlab{b}}), \eprint{0807.4909}.

\bibitem[Ambrosino et~al.(2009{\natexlab{c}})]{etappee}
F.~Ambrosino, et~al., \emph{Phys. Lett.} \textbf{B675}, 283--288
  (2009{\natexlab{c}}), \eprint{0812.4830}.

\bibitem[Landsberg(1985)]{landsberg}
L.~G. Landsberg, \emph{Phys. Rept.} \textbf{128}, 301--376 (1985).

\bibitem[Gao(2002)]{gao}
D.-N. Gao, \emph{Mod. Phys. Lett.} \textbf{A17}, 1583--1588 (2002),
  \eprint{hep-ph/0202002}.

\end{thebibliography}
\end{document}